\documentclass{emulateapj}

\usepackage{graphicx}
\usepackage{amsmath}
\usepackage{amssymb}

\newcommand{\Nifs}{\ensuremath{^{56}\mathrm{Ni}}}

\newcommand{\cms}{\ensuremath{{\rm cm~s}^{-1}}}
\newcommand{\kms}{\ensuremath{{\rm km~s}^{-1}}}
\newcommand{\ergss}{\ensuremath{{\rm ergs~s}^{-1}}}
\newcommand{\gcms}{\ensuremath{{\rm cm}^2{\rm g}^{-1}}}

\newcommand{\Msun}{\ensuremath{M_\odot}}
\newcommand{\Mni}{\ensuremath{M_{\rm ni}}}

\newcommand{\rd}{\ensuremath{r_{\rm d}}}

\newcommand{\ti}{\ensuremath{t_{\rm i}}}

\newcommand{\tsn}{\ensuremath{t_{\rm sn}}}

\newcommand{\tday}{\ensuremath{t_{\rm day}}}

\newcommand{\Rwd}{\ensuremath{R_{\rm wd}}}
\newcommand{\dr}{\ensuremath{l_{\rm sh}}}
\newcommand{\ldif}{\ensuremath{l_{\rm d}}}

\newcommand{\vt}{\ensuremath{v_{\rm t}}}
\newcommand{\vm}{\ensuremath{v_{\rm max}}}
\newcommand{\vmin}{\ensuremath{v_{\rm min}}}     

\newcommand{\omegah}{\ensuremath{\Omega_{\rm h}}}
\newcommand{\thetah}{\ensuremath{\theta_{\rm h}}}

\newcommand{\crho}{\ensuremath{\zeta_{\rho}}}
\newcommand{\cvt}{\ensuremath{\zeta_{v}}}
\newcommand{\rhos}{\ensuremath{\rho_{\rm s}}}
\newcommand{\rhof}{\ensuremath{\rho_{\rm f}}}
\newcommand{\pss}{\ensuremath{p_{\rm s}}}
\newcommand{\pff}{\ensuremath{p_{\rm f}}}
\newcommand{\Ts}{\ensuremath{T_{\rm s}}}

\newcommand{\kaps}{\ensuremath{\kappa_{e}}}
\newcommand{\Lprompt}{\ensuremath{L_{\rm x,iso}}}
\newcommand{\Eprompt}{\ensuremath{E_{\rm x}}}
\newcommand{\dtprompt}{\ensuremath{\Delta t_{\rm x}}}
\newcommand{\Tprompt}{\ensuremath{T_{\rm x}}}
\newcommand{\Lni}{\ensuremath{L_{\rm ni}}}
\newcommand{\Lopt}{\ensuremath{L_{\rm c,iso}}}

\newcommand{\ath}{\ensuremath{a_{\rm 13}}}
\newcommand{\Rs}{\ensuremath{R_\star}}
\newcommand{\Mc}{\ensuremath{M_{\rm c}}}

\shortauthors{D. Kasen}
\shorttitle{Supernova Companion Collision Emission}
\begin{document}

\title{Seeing the  Collision of a Supernova with its Companion Star}

\author{Daniel Kasen\altaffilmark{1,2}\email{kasen@ucolick.org}}

\altaffiltext{1}{University of California, Santa Cruz}
\altaffiltext{2}{Hubble Fellow}

\begin{abstract} 
  The progenitors of Type~Ia and some core collapse supernovae are
  thought to be stars in binary systems, but little observational
  evidence exists to confirm the hypothesis.  We suggest that the
  collision of the supernova ejecta with its companion star should
  produce detectable emission in the hours and days following the
  explosion. The interaction occurs at distances $\sim
  10^{11}-10^{13}$~cm and shocks the impacting supernova debris,
  dissipating kinetic energy and re-heating the gas.  Initially, some
  radiation may escape promptly through the evacuated region of the
  shadowcone, producing a bright X-ray (0.1-2~keV) burst lasting
  minutes to hours with luminosity $L \sim 10^{44}~\ergss$.
  Continuing radiative diffusion from deeper layers of shock heated
  ejecta produces a longer lasting optical/UV emission which exceeds
  the radioactively powered luminosity of the supernova for the first
  few days after the explosion.  These signatures are prominent for
  viewing angles looking down upon the shocked region, or about 10\%
  of the time.  The properties of the emission provide a
  straightforward measure of the separation distance between the stars
  and hence (assuming Roche lobe overflow) the companion's radius.
  Current optical and UV data sets likely already constrain red giant
  companions.  By systematically acquiring early time data for many
  supernovae, it should eventually be possible to empirically determine
  how the parameters of the progenitor system influence the outcome of
  the explosion.
\end{abstract}

\section{Introduction}

Observations of supernova light curves and spectra have allowed us to
characterize the outcome of the explosion -- the burned and ejected
stellar debris -- in remarkable detail.  But we still know very little
about the starting point.  In the most widely considered scenario,
Type Ia supernovae (SNe~Ia) result from carbon/oxygen white dwarfs
which reach a critical mass by accreting material from a
non-degenerate companion star.  Observational confirmation of the
binary nature of the progenitor system is lacking, however, and almost
nothing is known about the properties and diversity of the companion
stars.  The proposed progenitors should be rather dim, so it is not
surprising that we have so far failed to find them in pre-explosion
images of the host galaxies \citep[e.g.,][]{Maoz_companion}.  One
therefore seeks other means of inferring the presence of a stellar
companion.

A few minutes to hours after the supernova eruption, the debris
ejected in the explosion is expected to overrun the companion.  The
star is shocked by the impact, and it's envelope partially stripped
and ablated, but it survives the ordeal \citep{Wheeler_binary,
  Fryxell_Arnett,Chugai_binary, Livne_binary, Marietta, Pakmor}.
Observations of the remnant of Tycho's 1572 supernova turned up a
high-velocity G star, claimed to be the runaway companion
\citep{Ruiz_Lapuente_Tycho}.  This identification is still debated
\citep{Kerzendorf}.  Meanwhile, several attempts to look for evidence
of stripped hydrogen in the spectra of SNe~Ia have detected nothing
\citep{Mattila_01el, Leonard_stripped}.  The supernova ejecta is
distorted by the collision, which should lead to polarization of the
supernova light \citep{Kasen_hole}.  But although polarization has
been detected in several SNe~Ia \citep{Wang_ARAA}, there is nothing to
unambiguously indicate that this asymmetry relates to companion
interaction.

While the previous investigations have focused on the long term
consequences, one might wonder: could we see the collision itself?
Here we can draw an interesting parallel with core collapse supernova
explosions, in which a shock wave propagates through the envelope of a
massive star.  When that shock front nears the stellar surface at
radius $R \approx 10^{11}-10^{13}$~cm, the post shock energy vents in
an X-ray breakout burst lasting minutes to hours
\citep{Klein_Chevalier_1978,Matzner_Mckee}.  In the days that follow,
optical/UV radiation continues to diffuse from the deeper layers of
shock heated ejecta.  Eventually, the energy deposition from
radioactive \Nifs\ decay takes over.  The early shock luminosity phase,
however, has been observed in several events, e.g., SN~1987A
\citep{Arnett_1987A} and SN~1993J \citep{Wheeler_93J} while the shock
breakout itself was caught for SN~2008D \citep{Soderberg_08D,
  Modjaz_08D}.

The same physics applies to Type~Ia supernovae, but because the radius
of the progenitor white dwarf is so small ($\Rwd \approx 2 \times
10^8$~cm) the breakout emission should be extremely brief and the
early luminosity remarkably dim.  The problem is that when energy
input occurs at small radii, adiabatic losses in the rapidly expanding
($v \approx 10^9~\cms$) ejecta are overwhelming, and the thermal shock
deposited energy is converted to kinetic energy on the expansion timescale
($\Rwd/v \sim 0.1$~sec), which is much shorter than the diffusion
timescale.

As it turns out, the separation distance between the white dwarf and
its companion star is presumed to be $a \sim 10^{11}-10^{13}$~cm,
comparable to the radii of core collapse progenitors.  When the
supernova ejecta collides with its companion, the impacting layers are
re-shocked and the kinetic energy partially dissipated.  If the geometry
is favorable, some of this energy might escape straightaway in a
prompt burst, which will be followed by a longer-lasting tail of
diffusive emission -- the analogues of shock breakout and its
aftermath in core collapse events.  In this case the emission provides
a measure not of the stellar radius $\Rwd$, but of the separation
distance $a$.

Here we develop an analytic description of the collision dynamics
and subsequent radiation transport which suggests that early time
observations of supernovae at X-ray through optical wavelengths should
offer a powerful means of confirming the presence of a companion star
and constraining its parameters.

\section{Collision Dynamics }
\label{sec:collide}

\begin{figure}
\includegraphics[width=3.4in]{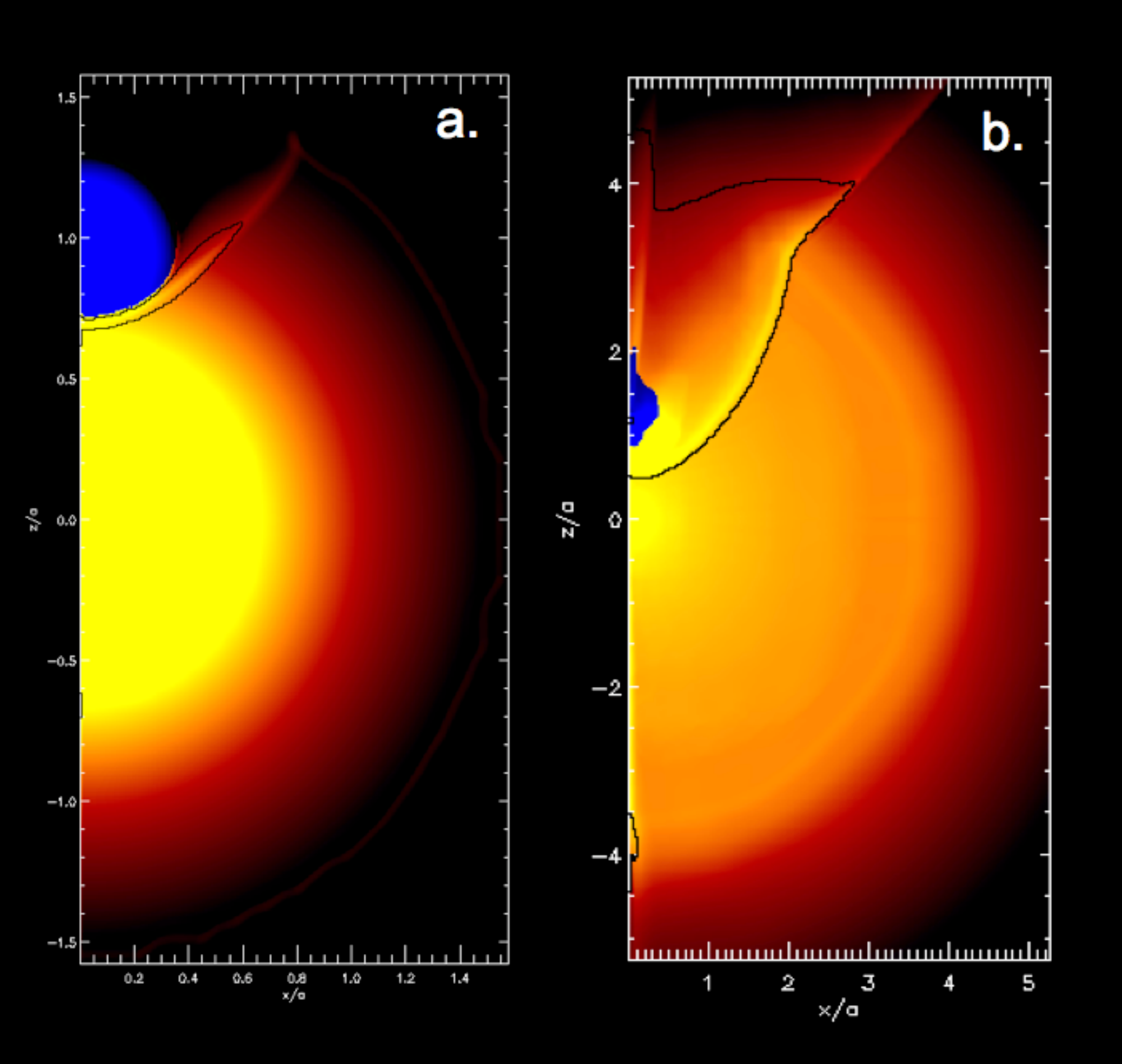}
\caption{Hydrodynamical calculation of a Type~Ia supernova colliding with a red giant star ($\Rs = 10^{13}$~cm, $a = 2.5 \times 10^{13}$~cm) {\bf a:} Density structure during the prompt emission phase ($t = \ti/2$).  The companion star (drawn in blue) diverts the flow and carves a hole in the ejecta.   The black contour shows the region where the shocked ejecta temperature exceeds $3\times 10^5$~K. {\bf b:}  Density structure at a later phase ($t = 2 \ti$, note change of scale).  The shell of shocked ejecta has expanded to partially refill the hole.  The black contour shows the region where the temperature exceeds $10^5$~K.
  \label{Fig:collide}}~~~~~~~~~~~~~~~
\end{figure}

In the single degenerate scenario of SNe~Ia, the companion stars are
thought to be either slightly evolved main sequence stars or red
giants \citep{Branch_companions, Hachisu_RG}.  For the most promptly
exploding systems, the companions may be 5-6~\Msun\ main sequence (MS)
sub-giants, with radii $\Rs \sim 5 \times 10^{11}$~cm.  More commonly,
the MS companions may be $1-3~\Msun$ sub-giants with radii $\sim 1-3
\times 10^{11}$~cm.  In the red giant (RG) case, the stars are evolved
$\sim 1-2~\Msun$ stars with $\Rs \sim 10^{13}$~cm.  In most scenarios,
the companion is believed to be in Roche lobe overflow.  The
separation distance, $a$, is then comparable to the companion radius;
for typical mass ratios, $a/\Rs = 2-3$.

After the supernova explodes, the ejected debris expands freely for
some time before hitting the companion.  The flow becomes homologous,
and the radius of a fluid element $r = vt$ where $v$ is the velocity
and $t$ is the time since explosion.  We will describe the ejecta
density profile by a broken power law \citep{Chevalier_Soker} with a
shallow inner region $\rho_{\rm i} \propto r^{-\delta}$ and a steep outer
region $\rho_0 \propto r^{-n}$.  The profiles join at the transition
velocity
\begin{equation}
\vt = 6 \times 10^8~\cvt (E_{51}/\Mc)^{1/2}~\cms,
\end{equation}
where $E_{51} = E/10^{51}$~ergs is the explosion energy, $\Mc =
M/M_{\rm ch}$ is the ejecta mass in units of the Chandrasekhar mass,
and \cvt\ is a numerical constant. The density in the outer layers ($v
> \vt$) is
\begin{equation}
\rho_0(r,t) = 
\crho \frac{M}{\vt^3 t^3}  \biggl(\frac{r}{  \vt t } \biggr)^{-n}
\end{equation}
with a similar expression for the inner layers.  The numerical
constants follow from the requirement that the density integrate to
the specified mass and kinetic energy
\begin{equation}
\begin{split}
\cvt &= \biggl[ \frac{2(5-\delta)(n-5)} {(3-\delta)(n-3)} \biggr]^{1/2} \\
\crho &=  \frac{1}{4 \pi} \frac{(n-3)(3-\delta)}{n-\delta}.
\end{split}
\end{equation}
The broken power law profile was originally derived for core collapse
supernovae, but it fits multi-dimensional delayed-detonation models of
SNe~Ia remarkably well.  For the models of \cite{Kasen_WLR} we find
typical values of $\delta=1$, $n=10$, in which case the constants are
$\cvt = 1.69$, $\crho = 0.12$.

The characteristic timescale for the supernova ejecta to interact with the companion is
\begin{equation}
\ti = a/\vt =  10^4~\ath v_9^{-1}~{\rm sec},
\end{equation}
where $\ath = a/10^{13}$~cm, and $v_9 = \vt/10^9~\cms$.  For RG
companions the interaction timescale $\ti \approx 5$~hours, while for
MS sub-giants, $\ti \approx 5-30$~minutes.

The ejecta is highly supersonic when it collides with the companion,
with mach number $\mathcal{M} \approx (a/\Rwd)^{1/2} \gg 1$.  Figure~1
illustrates the hydrodynamics of the interaction in a 2-dimensional
numerical calculation using the FLASH code \citep{Fryxell_FLASH} and
assuming a polytropic $\gamma = 4/3$ equation of state, appropriate
for a radiation dominated gas.   As the flow sweeps over the companion
star, a bow shock forms.  Ejecta passing through the shock is heated
and compressed into a thin shell, and its velocity vector is
redirected.

We will rely here on an approximate analytic description of the
collision dynamics.  Material moving at velocity $v$ interacts at a
time $t_v \approx a/v$.  The ejecta properties immediately
after being shocked are given by the Rankine-Hugoniot jump conditions
in the hypersonic limit.  The density of the shocked gas is
\begin{equation}
\rhos(v) = \frac{\gamma + 1}{\gamma -1} \rho_0(a,t_v) 
= 7 \crho \frac{M}{a^3} \biggl( \frac{v}{\vt} \biggr)^{-n+3}
\label{Eq:rhos}
\end{equation}
taking $\gamma = 4/3$.  The pressure
of the shocked gas is of order the incoming ram pressure
\begin{equation}
\pss(v) = \frac{2}{1 + \gamma}  ~\rho_0 v^2 \sin^2 \chi
= \frac{6}{7} \crho \sin^2\chi \frac{M \vt^2}{a^3}  \biggl( \frac{v}{\vt} \biggr)^{-n+5}
\label{Eq:ps}
\end{equation}
where $\chi$ is the angle of the oblique shock front relative to the
flow direction.  The actual value of $\chi$ varies along the bow
shock, but for simplicity we take a constant, characteristic value
near the maximal turning angle for hypersonic flows, $\chi \approx
45^\circ$.

For a radiation dominated gas, $\pss = a_R \Ts ^4/3$, where $a_R$ is
the radiation constant, which gives the equilibrium temperature of the shocked
debris 
\begin{equation} \Ts(v) = 2.8 \times 10^6~\Mc^{1/4} v_9 ^{1/2}
  \ath^{-3/4} \bigg( \frac{v}{\vt} \biggr)^{-(n-5)/4}~{\rm K}
  \label{Eq:Ts} \end{equation} 
For a RG companion at $a = 2.5\times
10^{13}$~cm, the outer layers of ejecta ($v \sim 3 \vt$) have $\Ts
\approx 3\times 10^5$~K, a value confirmed by the numerical
calculation (Figure~1a).  MS companions will have higher shock
temperatures with $\Ts \approx 10^6-10^7$~K.  
 To check the assumption of
radiation domination in the shocked region, we note that   Eqs.~\ref{Eq:rhos} and \ref{Eq:ps}
imply a ratio of radiation energy to 
 electron/ion energy
\begin{equation}
\frac{ a \Ts^4}{\rhos k_B \Ts/m_p} =  707~\ath^{3/4} \Mc^{-1/4} v_9^{3/2}
\end{equation}
which is $\gg 1$ for the scenarios under consideration.  The details of
thermalization, however, deserve further investigation.  Initially,
the electrons/ions are shocked to very high temperatures ($\sim 10^9-10^{10}$~K),
then radiatively cool toward the equilibrium value Eq.~\ref{Eq:Ts} by processes similar to those in
ordinary supernova shocks -- i.e., bremsstrahlung followed by Compton
up-scattering, and in some cases pair-production  \citep{Weaver_shock}.  If 
the timescale for the gas to radiate lags the dynamical timescale, 
non-equilibrium temperatures significantly
greater than $\Ts$  can be realized in the relaxation region.   This is likely to occur
in the highest-velocity, lowest-density outermost layers, and may be a source
of harder radiation \citep{Katz_shock}.

The interaction with the companion diverts the incoming flow, carving
out a conical hole in the supernova ejecta.  The half opening angle of
the hole is roughly $\thetah = \tan^{-1} (r_{\rm b}/a)$ where $r_{\rm b}$ is the
extant of the bow shock.  Simulations find $r_{\rm b} \sim 2 \Rs$ and
$\thetah = 30-40^\circ$ (see Figure~1 and \cite{Marietta}). The solid
angle, \omegah, of the hole is
\begin{equation}
\frac{\omegah}{4\pi} = \frac{1}{2} (1 - \cos \thetah)
 = \frac{1}{2} \biggl[ 1 - \frac{1}{1 + (r_{\rm b}/a)^2} \biggr] \approx \frac{1}{10}
\end{equation}
This hole will provide a channel for radiation to quickly escape from
the otherwise optically thick ejecta.

The ejecta displaced from the hole piles up into a compressed shell
along the cone surface.  Assuming this shell layer is thin, its thickness
\dr\ can be estimated by mass conservation.  The volume of a region of
radial extent $dr$ within the conical cavity is $V_{\rm i} = \omegah a^2
dr$.  The gas swept out of this region occupies a volume $V_{\rm f} = 2 \pi
a \dr dr$.  The condition $\rho_0 V_{\rm i} = \rhos V_{\rm f}$ gives
\begin{equation}
\frac{\dr}{a} = \frac{\omegah}{4 \pi} \frac{2 \rho_0}{\rhos}  \approx \frac{1}{35}
\label{Eq:lsh}
\end{equation}
A dense shell of roughly this thickness is seen in Figure~1a.  The actual dynamics can become
quite complex, with the shell broken in pieces by shear instabilities and the companion
envelope shredded.

After passing by the companion star, the shocked gas can expand
laterally to try to refill the evacuated hole.  The situation
resembles the isentropic expansion of a gas cloud into vacuum
\citep{Zeldovich_Raizer}; the front of the rarefaction wave moves
outward at the maximum escape velocity $v_l = 2/(\gamma-1) c_s$ where
$c_s = (\gamma \pss/\rhos )^{1/2} = (8/49)^{1/2} \sin \chi~v$ is the sound speed of the shocked
material.  The net velocity in the direction perpendicular to the
symmetry axis is then $v_x = v \sin \thetah - v_l \cos \thetah \approx
-0.2~v$.  The ejecta moving at velocity $v$ thus re-closes at a time
$\Rs/|v_x| \approx 5 \Rs/v$ after passing by the companion, or at a time
$t_h = (a + 6 \Rs)/v$ after the explosion.   
 In the adiabatic calculation
 (Figure~1b), the rarefaction softens
the density gradient in the polar direction, but fails to
refill the shadowcone region uniformly before freezing out.  
If  radiative cooling is significant during these phases (see Section~\ref{sec:xray}) 
the sound speed  will be reduced, which would further delay or halt the closing of the hole. 

The energy density of the shocked gas is $\epsilon_{\rm s} = 3 \pss$ and so
the total energy dissipated in the collision shock is found to be
\begin{equation}
E_{\rm th} = \frac{18}{49}  \sin^2\chi \frac{\omegah}{4 \pi} E \approx
1.5 \times 10^{49}~E_{51}~{\rm ergs}.
\end{equation}
Much of this thermal energy will be lost again to adiabatic expansion,
but if even a fraction is radiated the collision luminosity should be
quite bright.

\section{Prompt X-ray Burst}
\label{sec:xray}

As supernova ejecta flows past the companion star, the hot surface
layers of the shocked shell become exposed (Figure~1a). At this time, some radiation may be able
to escape straightaway through the evacuated shadowcone hole, giving rise to a
sudden burst of emission.
In general, only a fraction of the total energy in the  shell
can be radiated promptly -- i.e., before suffering significant loses
due to adiabatic expansion.  This prompt emission arises from the surface
layer of the shell with a thickness, $\ldif$, determined by requiring
that the diffusion time through that layer
\begin{equation}
t_d = \tau \frac{\ldif}{3 c} = \frac{\ldif^2 \kappa \rhos}{ 3c} 
\end{equation}
be less than the timescale for expansion, given by the shell sound crossing time $\dr/c_s$
(which is typically shorter than the dynamical timescale $a/v$).  
Using Eqs.~\ref{Eq:rhos} and \ref{Eq:lsh} gives
 \begin{equation}
\frac{\ldif}{\dr} = 
\frac{a}{\vt \tsn} 
\biggl( \frac{4 \pi}{\omegah} \biggr)^{1/2}
\biggl( \frac{\sqrt{8}}{14 \crho \sin \chi}  \biggr)^{1/2}
\biggl[\frac{v}{\vt} \biggr]^{(n-4)/2}
\end{equation}
where the quantity
\begin{equation}
\tsn = \biggl( \frac{\kappa M}{3 c \vt} \biggr)^{1/2} = 29~\Mc^{1/2} v_9^{-1/2} \kaps^{1/2}~{\rm days}
\end{equation}
is the familiar "effective" diffusion time \citep{Arnett_1982} that
sets the duration of the ordinary \Nifs-powered SN~Ia light curve.  We
have assumed a constant opacity $\kaps = 0.2~\gcms$, appropriate for
electron scattering in fully ionized $A/Z = 2$ elements.  For RG
companions at $a \approx 10^{13}$~cm, we find $\ldif \simeq \dr$ for the
layers $v \ga 2 \vt$.  In this case, most of the energy dissipated in the outers layers
can be radiated in the prompt burst.  For MS
companions with $a \simeq 10^{11}-10^{12}$~cm, the ratio $\ldif/\dr \simeq 0.01-0.1$, and
only a fraction of the photons escape promptly.

While the bulk of the SN ejecta remains extremely optically thick at
this phase, photons can initially escape through the hole carved out
in the interaction.  This channel will close, however, once the outer
layers of ejecta have re-expanded to fill the shadowcone, which
happens at a time $t_h \approx (a + 6 \Rs)/\vm$ where \vm\ is the maximum
ejecta velocity (see Section~\ref{sec:collide}).  A given layer
of ejecta can contribute to the burst only if it passes the companion
at a time less than $t_h$, which holds for material moving
faster than $\vmin \simeq \vm(a + \Rs)/(a + 6 \Rs)$.  For $a/\Rs = 3$,
we find $\vmin \approx \vm/2$.

By integrating the energy density, $\epsilon_{\rm s} = 3 \pss$, of the shocked
gas within \ldif,
we can estimate the total energy escaping in the burst
\begin{equation}
\begin{split}
\Eprompt =& 
~  2\pi \int_{\vmin}^{\vm} 3 \pss(v) \ldif(v) ~a \biggl( \frac{a}{v} \biggr) dv  \\
 =&~ 7.72 \times 10^{47}~
 \biggr(\frac{\omegah}{4 \pi} \crho \sin^3 \chi \biggl)^{1/2}~
 \biggr(\frac{62}{n-6} \biggl)
 \times \\
 & a_{13} \Mc^{1/2} v_9^{3/2} \kaps^{-1/2} 
\biggl( \frac{\vt}{\vmin} \biggr)^{(n-6)/2}
 ~{\rm ergs},
\end{split}
\end{equation}
where we approximated the upper limit as $\vm \rightarrow \infty$.
The duration of the burst is the time it takes the emitting ejecta to
flow past the companion, or $\dtprompt = (a +\Rs)/\vmin - (a + \Rs)/\vm$.  
For typical values ($\vmin \approx 2 \vt$; $\vm \approx 4 \vt$; $\Rs \approx 3 a$) 
this duration is $\dtprompt \approx \ti/3$.  Assuming the radiation is
emitted into a solid angle \omegah, the isotropic equivalent
luminosity is $\Lprompt = (4\pi/\omegah) (\Eprompt/\dtprompt)$.  Taking
characteristic values $(n,\delta,\chi,\thetah,\vmin) =
(10,1,45^\circ,40^\circ,2 \vt)$ we find
\begin{equation}
 \begin{split}
\Lprompt = 
5.8 \times 10^{44}~
  \Mc^{1/2} v_9^{5/2} \kaps^{-1/2}
  ~\ergss
 \end{split}
 \end{equation}
This luminosity is independent of $a$, and so
 roughly comparable
for all types of companions.
The value is similar to
 that of shock breakout in core collapse SNe, which is not surprising  given
 that the  shock temperature and emitting surface area  are comparable in the two phenomena.
 The collision burst will only be visible for viewing  angles peering down the hole, $\theta \la \thetah$.  
Such an  orientation occurs $\omegah/4 \pi \approx 10\%$ of the time.  

 The spectrum of the prompt burst may be approximated as a blackbody at the equilibrium shock
 temperature (Eq.~\ref{Eq:Ts}), implying emission peaking in the soft X-ray
 with typical energies $\Tprompt \sim 0.05-0.1$~keV for RG and $\Tprompt \sim 0.2-2$~keV
 for MS companions. 
 Non-equilibrium effects (see Section \ref{sec:collide})  could lead to some emission at
significantly higher energies ($10-100$~keV), while non-thermal
particle acceleration may also contribute a power law continuum of hard radiation.
X-rays emitted in the direction of the companion star will ionize its surface layers and 
be reprocessed,
likely giving rise to 
 substantial line recombination/fluorescence emission 
\citep[e.g.,][]{Ballantyne_2001}.  

 According to Eq.~\ref{Eq:Ts} the temperature of the burst spectrum evolves 
 with time as
   \begin{equation}
   T_x(t) = 0.1~ \ath^{-3/4}  \Mc^{1/4} v_9 ^{1/2}
 \bigg( \frac{t}{\ti/2} \biggr)^{(n-5)/4}~{\rm keV},
  \label{Eq:Tx} \end{equation} 
which predicts that, at least initially, the spectrum becomes harder
with time, as interior layers of ejecta have higher densities and
shock temperatures.  On the other hand, deviations from equilibrium
are likely to be greatest in the highest velocity layers, which may
counteract this trend.  Eventually, as ejecta gradually refills the
shadowcone, the effective photosphere moves to a larger radius and
the emission must soften.

The actual structure of the collision region will be more complex and
inhomogeneous than that imagined here, due either to the inherent
clumpiness of the supernova ejecta, or to secondary shocks and
hydrodynamical instabilities developing in the interaction
\citep[e.g.,][]{Cid-Fernades}.  This may lead to fluctuations in the
burst light curve on a time scale $\delta R/v$ where $\delta R$ is the
typical clump size.  Multi-dimensional radiation-hydrodynamics
calculations will be needed to characterize the light curve and
spectra in detail.

\section{Early UV/Optical Luminosity}

\begin{figure}
\includegraphics[width=3.5in]{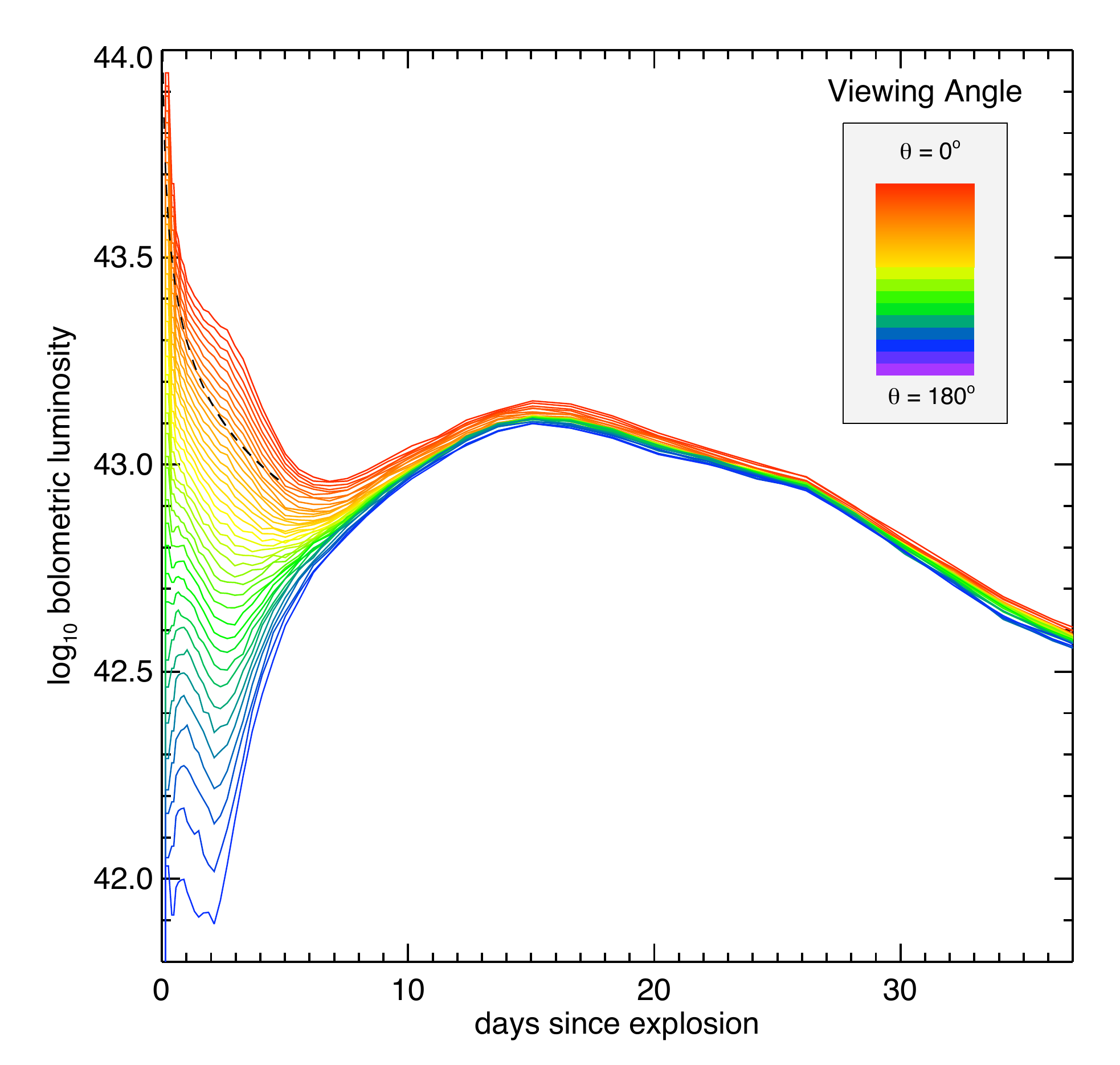}
\caption{Model light curve of a Type~Ia supernova having collided with a red giant companion at a separation distance $a = 2 \times 10^{13}$~cm.  The luminosity due to the collision is prominent at times  $t< 8$~days and for viewing angles looking down on the collision region  ($\theta = 0^\circ$).   At later times the emission is powered by the radioactive decay of 0.6~\Msun\ of \Nifs\ located in the inner layers of ejecta ($v < 10^9~\cms$).
The black dashed line shows the analytic prediction for the early phase luminosity (Eq.~\ref{Eq:Lc}). ~~~~~~~~~~~~~~~~
 \label{Fig:bol}}

\end{figure}

Thermal energy not radiated in the prompt burst can diffuse out in the
hours and days that follow, but will suffer loses from adiabatic
expansion.  At $t \sim 1$~day, this emission will be primarily at
UV/optical wavelengths.

We consider here times $\gg \ti$ such that homology has been
re-established in the ejecta.  The final ejecta structure will be
asymmetric, but for now we neglect angular dependencies.  The density
profile is taken to be $\rhof(r,t) = 7 f_0~\rho_0(r,t)$, where the
constant $f_0 < 1$ accounts for the lateral expansion and radial
readjustment that occur during the transition to homology.

Assuming the evolution is adiabatic, the pressure profile of a
shocked fluid element at these times is $\pff(t) = \pss [\rhof(t)/\rhos]^\gamma$.
For the region affected by the collision ($\theta < \thetah$)
\begin{equation}
\pff(r,t) = \frac{6}{7} f_0^{4/3} \crho \sin^2 \chi \frac{M a}{\vt^2 t^4}  \biggl( \frac{r}{\vt t} \biggr)^{-n+1}
\label{Eq:pf}
 \end{equation}
 The pressure is negligible outside \thetah.  As time progresses, the
 adiabatic profile Eq.~\ref{Eq:pf} will continue to describe the
 opaque inner layers of ejecta, but the outer layers will be modified
 by radiative diffusion.  The evolution can be calculated using a
 self-similar diffusion wave analysis \citep{Chevalier_1992}.  At a
 time $t$, diffusion will have affected the ejecta above a radius
 $\rd$ determined by setting the diffusion time $t_d = \rd^2 \kappa
 \rhof/3 c$ equal to the elapsed time $t$
\begin{equation}
\rd(t) = \biggl[ \frac{\crho}{3} \tsn^2 \biggr]^{1/(n-2)} \vt t^{(n-4)/(n-2)}
\label{Eq:rd}
\end{equation}
Over time, the diffusion wave recedes into the ejecta in a Lagrangian
sense.  However, for times $t \la 5$~days, $\rd$ remains in the steep
outer layers of ejecta ($v > \vt$).

Considering only the radial transport, the isotropic equivalent
luminosity in the comoving frame is given by the diffusion
approximation
\begin{equation}
\Lopt(r,t) = 
-4 \pi r^2 \frac{c}{\kappa \rhof} \frac{\partial \pff}{\partial r}
\label{Eq:Ld}
\end{equation}
We continue to assume a constant opacity, even though at UV
wavelengths the line expansion opacity may exceed electron scattering
and will, to some extant, be temperature dependent.

In the outer layers of ejecta ($r > \rd$) the comoving luminosity
$\Lopt$ is constant with radius.  Its value is therefore set by
processes near the diffusion wave radius.  A reasonable estimate of
$\Lopt$ is derived by evaluating Eq.~\ref{Eq:Ld} at $\rd$, using the
pressure profile Eq.~\ref{Eq:pf},
\begin{equation}
\begin{split}
\Lopt = \frac{16 \pi}{49} (n-1) \eta \sin^2\chi  f_0^{1/3} \biggl[ \frac{\crho }{3} \biggr]^{2/(n-2)}\\ 
\times  \biggl( \frac{\ti}{\tsn^2} \biggr) \frac{1}{2} M \vt^2 \biggl( \frac{t}{\tsn} \biggr)^{-4/(n-2)}
\end{split}
\end{equation}
where $\eta$ is a constant of order unity that must be determined by
solving the full diffusion equation \citep{Chevalier_1992}.  Written
in this form, the resemblance to core collapse supernova light curves
is clear: apart from constants, the luminosity is of order $E/\tsn$
times a factor $\ti/\tsn$ that accounts for adiabatic loses.

Taking $(n,\delta,\chi,f_0,\eta) = (10,1,45^\circ,0.5,0.5)$ we find that from appropriate viewing angles
\begin{equation}
\Lopt =   10^{43}~ \ath~\Mc^{1/4} v_9^{7/4} \kaps^{-3/4} \tday^{-1/2}~\ergss,
\label{Eq:Lc}
\end{equation}
where \tday\ is the time since explosion measured in days.  The
derivation applies only at times significantly greater than $\ti$, but
Eq.~\ref{Eq:Lc} may provide a workable estimate at earlier times.  The
observer frame luminosity differs from Eq.~\ref{Eq:Lc} by an
additional term accounting for the advected luminosity, of order $v/c
\la 10\%$ compared to $\Lopt$.

The collision luminosity will only be readily discernible when it
exceeds the luminosity, \Lni,  of the ordinary \Nifs\ powered light curve.  At
early times, $t \ll \tsn$, the approximate analytic light curves of
SNe~Ia give $\Lni = \epsilon_{\rm ni} M_{\rm ni} (t/\tsn)^2$, where
$\epsilon_{\rm ni} = 4.8\times 10^{10}$~ergs~s$^{-1}$g$^{-1}$ and
$M_{\rm ni}$ is the mass of \Nifs\ \citep{Arnett_1982}.  We then find
that $\Lopt > \Lni$ for times
\begin{equation}
t_{\rm c} < 7.3~  \ath^{2/5} \Mc^{1/2} v_9^{3/10} \kaps^{1/10}
\biggl( \frac{\kappa_{\rm ni}}{\kaps} \biggr)^{2/5}
M_{{\rm ni},0.6}^{-2/5}
~{\rm days},
\end{equation}
where $M_{\rm ni,0.6} = \Mni/0.6\Msun$. Note that the opacity in the
\Nifs\ region, $\kappa_{\rm ni}$, is heavily affected by iron group
line blanketing, and so may be greater than the opacity \kaps\ in the
outer layers.  This would help delay the \Nifs\ luminosity takeover.

The wavelength of the early emission depends on the photospheric
radius, $r_p$, defined as the location where the optical depth $\tau =
\int_{r_p}^\infty \rhof \kappa dr = 1$, or
\begin{equation}
r_p =  t^{(n- 3 )/(n-1)} \biggl( \frac{\crho \kappa M \vt^{(n-3)} }{n-1} \biggr)^{1/(n-1)}
\end{equation}
The effective temperature of the emission $T_{\rm eff} = ( \Lopt/4 \pi r_p^2 \sigma)^{1/4}$ is then, for $n=10$
\begin{equation}
T_{\rm eff} = 2.5 \times 10^4~\ath^{1/4} \kaps^{-35/36}~ 
\tday^{-37/72}~{\rm K}.
\end{equation}
At $t = 1$~day, the emission peaks at wavelengths $\lambda \approx
1000$~\AA, however the Rayleigh-Jeans tail of the blackbody extends
into the near UV and optical.

While the analytic solution captures the essential physical effects,
it neglects the ejecta asymmetry and non-radial transport.  We have
therefore calculated numerical light curves using a 3-D non-grey
radiation transfer code which also accounts for the effects of line
opacity and \Nifs\ heating \citep{Kasen_MC}.  For simplicity, 
we use an artificial
ejecta model with density profile $\rhof(r,\theta,t) = \rho_0(r,t)
f_0(\theta)$, where the function $f_0(\theta)$ now describes the
angular structure of the conical hole region
\begin{equation}
f_0(\theta) = f_h + (1 - f_h) \frac{x^m}{1 + x^m}
\biggl( 1 + A \exp
\biggl[-
\frac{(x-1)^2}{(\theta_p/\thetah)^2} 
\biggr] \biggr)
\end{equation}
where $x = \theta/\thetah$.  This formula approximates the results
of simulation in the homologous phase choosing $m=8$, $f_h = 0.1$. $\thetah = 30^\circ$,
$\theta_p = 15^\circ$, and $A = 1.8$.  The pressure profile in the
shocked region was taken from Eq.~\ref{Eq:pf}.

The light curve calculation (Figure~2) shows that the early collision
luminosity is dramatic when the companion is a RG at $a = 2 \times
10^{13}$~cm.  The numerical results agree with the analytic
estimate (Eq.~\ref{Eq:Lc}), and also illustrate the anisotropy of the
radiation.  The collision luminosity is brightest for viewing angles
looking down upon the shocked region ($\theta < \thetah$), but a
significant amount of radiation diffuses out at angles $\theta \approx
90^\circ$, and a few percent is even back-scattered along $\theta
\approx 180^\circ$.

While the companion interaction produces a conspicuous signature (a kink) in 
the early bolometric light
curve, most of the emission is in the UV; in the optical bands, the
effect is less dramatic (Figure~3).  For a RG companion, the B-band
light curve shows a distinct bump at $t < 5$~days which should be
relatively easy to probe observationally.  At redder wavelengths, or
for smaller separation distance $a$, the collision simply modifies the
shape of the light curve rise.  However, because SN~Ia light curves
are quite standard, a statistical analysis of (good quality) early
time photometry should be able to pull out these subtle differences.

\begin{figure}
\includegraphics[width=3.5in]{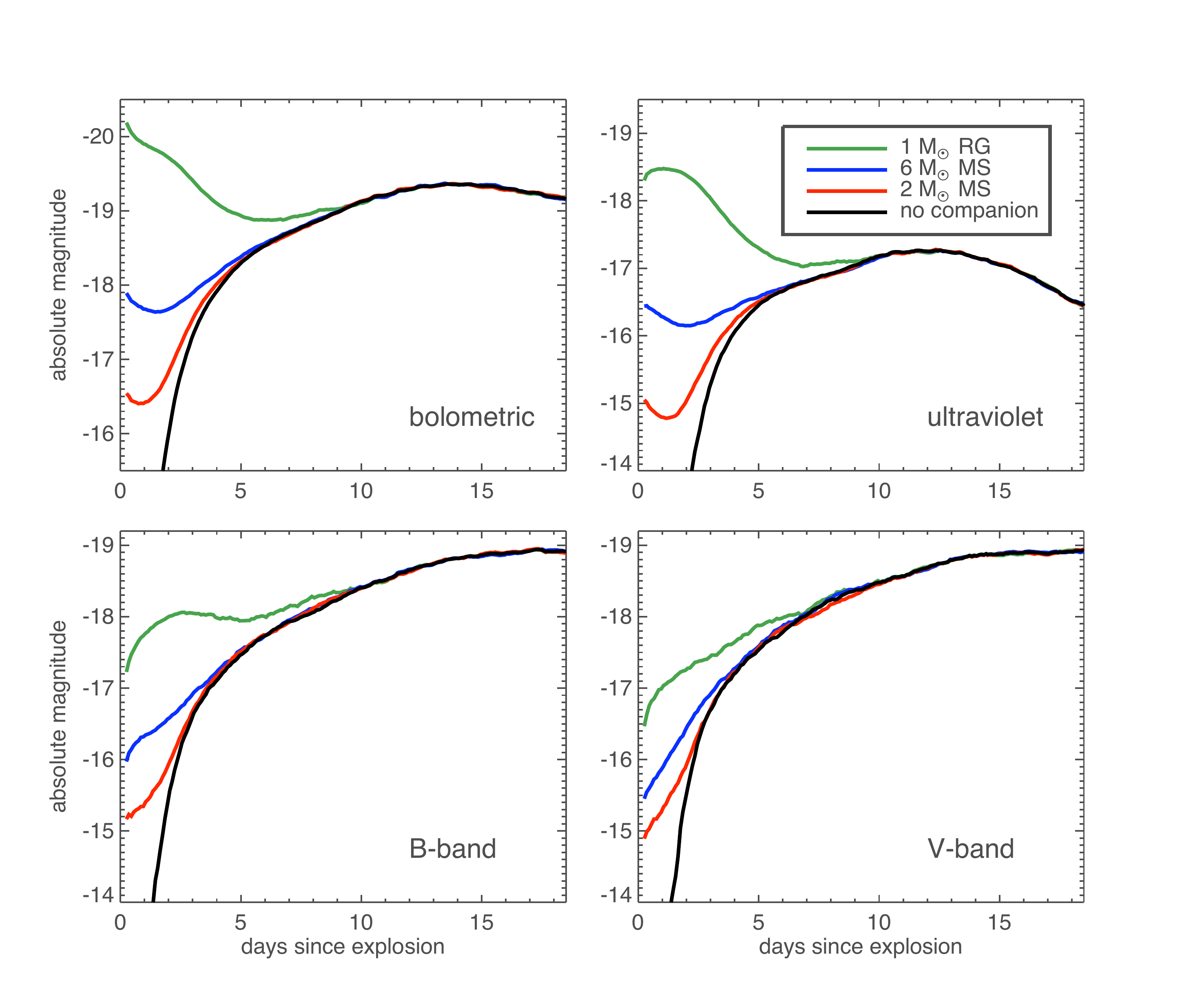}
\caption{
Signatures of companion interaction in the early broadband light
curves of Type~Ia supernovae.  We model three different progenitor
scenarios: a RG companion at $a = 2 \times 10^{13}$~cm (green lines),
a $6~\Msun$ MS companion at $a = 2 \times 10^{12}$~cm (blue lines) and
a $2~\Msun$ MS companion at $a = 5 \times 10^{11}$~cm (red lines).
The ultraviolet light curves are constructed by integrating the flux
in the region $1000-3000$~\AA\ and converting to the AB magnitude
system.  For all light curves shown, the viewing angle is $\theta =
0^\circ$.
\label{Fig:bb}}
\end{figure}

\section{Observational Prospects}

\begin{deluxetable*}{ccccccccc}[t]
\tablewidth{6.0in}
\tablecaption{Properties of Type~Ia Supernova Collision Emission -  Analytic Estimates}
\tablehead
{
\colhead{Companion}                & 
\colhead{a}                        & 
\colhead{\Eprompt}                & 
\colhead{\dtprompt}                & 
\colhead{\Lprompt}                & 
\colhead{\Tprompt}                & 
\colhead{\Lopt\  (1 day)}                & 
\colhead{$t_{\rm c}$}
}            
\startdata

RG ($M \sim 1~\Msun$)    &         $2 \times 10^{13}$~cm  &  $3.9 \times 10^{47}$~ergs &$1.9$ hours &$5.8 \times 10^{44}$   
&   $0.07$~keV  &  $2 \times 10^{43}$ & $9.6$~days \\
MS ($M \sim 6~\Msun$)      &         $2 \times 10^{12}$~cm  &  $3.9 \times 10^{46}$~ergs &$11$ mins &$5.8 \times 10^{44}$   
&   $0.2$~keV  &  $2 \times 10^{42}$ & $3.8$~days  \\
MS ($M \sim 2~\Msun$)      &         $5 \times 10^{11}$~cm  &  $9.6 \times 10^{45}$~ergs &$2.8$ mins &$5.8 \times 10^{44}$   
&   $1.0$~keV  &  $5 \times 10^{41}$ & $2.2$~days \\ 
MS ($M \sim 1~\Msun$)      &         $3 \times 10^{11}$~cm  &  $5.8 \times 10^{45}$~ergs &$1.7$ mins &$5.8 \times 10^{44}$   
&   $1.4$~keV  &  $3 \times 10^{41}$ & $1.8$~days

\enddata
\label{tab:models}
\end{deluxetable*}

The results derived here suggest a new means for constraining
supernova companions using early photometric observations.  Table~1
summarizes the analytic estimates of the collision emission for
various SNe~Ia progenitors.  The theoretically predicted signatures
appear to be quite robust, as they rely only on established physics
familiar from the core collapse SNe context.  However, further numerical studies
using multi-dimensional radiation-hydrodynamical calculations 
(and including non-equilibrium effects)
will be needed to refine the detailed light
curve and spectrum predictions.

 For all companion types, signatures of the collision will be
 prominent only for viewing angles looking down upon the shocked
 region, or $\sim 10\%$ of the time.  Detection will therefore require
 high cadence observations of many supernovae at the earliest phases
 ($\la 5$~days) and at the bluest wavelengths possible.  Ironically,
 these observations may sometimes be easier for distant SNe. At
 redshifts $z \ga 0.5$, the UV flux would be redshifted into the
 U-band, while cosmological time dilation would prolong the light
 curve by a factor $(1+z)$.

Detecting the collision signatures becomes significantly easier for
larger separation distances.  Current optical and UV data sets likely
already constrain red giant companions ($a \simeq 10^{13}$~cm).
Ongoing or upcoming surveys could be tuned to probe the larger ($M \ga
3$~\Msun) main sequence companions ($a \simeq 10^{12}$~cm).  Optical
detection of the smallest $\sim 1~\Msun$ main sequence companions ($a
\simeq 10^{11}$~cm) will be challenging, requiring measurement of
subtle differences in the light curves at $t \la 2$~day.  However in
all cases the prompt X-ray burst should be bright.  Proposed X-ray
surveys \citep[e.g., EXIST,][] {Grindlay_2005} may then detect a large
number of collision bursts every year, at least in the case of MS
companions which produce harder radiation.  If non-equilibrium or
non-thermal shock effects are operative, some hard radiation may
accompany all bursts.

The most compelling reason for studying the collision emission is that
it offers a straight-forward measure of the separation distance
between the stars.  This value can be determined from the duration of
the X-ray burst ($\dtprompt \simeq a/v$), or its temperature
(Eq.~\ref{Eq:Ts}), or from the luminosity of the early optical/UV
emission (Eq.~\ref{Eq:Lc}).  If we in turn assume that
the companion is in Roche lobe overflow, it's radius can be inferred
$\Rs \sim a/3 - a/2$.  In principle, the ratio $a/\Rs$ itself could be
constrained using a statistical sample of SNe~Ia, as the anisotropy of
the emission depends on the opening angle, \thetah, of the shocked
region of ejecta.

While our discussion has focused on Type~Ia supernovae, similar
signatures of companion interaction should apply to Type~Ib/Ic and
some Type~II~SNe which may arise from close massive binaries
\citep{Podsi_Ibc}.  In these cases, the shock breakout from the
exploding star will contribute to the luminosity on comparable
timescales, likely producing two peaks in the X-ray emission.  Some
gamma-ray bursts might also come from binary systems, and the
interaction of the relativistic material with a stellar companion
may produce another type of X-ray flare \citep{MacGRB}.

It is also possible that early emission from SNe
stems from a  collision not with the companion star, but with a
surrounding circumstellar medium (CSM).  To substantially decelerate
the ejecta, the CSM would need to have a mass $\sim 0.01-0.1~\Msun$ located at
radii $\sim 10^{11}-10^{13}$~cm.  A slow ($10~\kms$) stellar wind
moves beyond these distances is less than a year, so it may be
difficult to realize these conditions in a single degenerate scenario of
SNe~Ia.
In the double degenerate merger scenario, the total mass of the system
can exceed $M_{\rm ch}$, and a few 0.1~\Msun\ of excess
carbon/oxygen may linger in the vicinity.  If this material remains at
the tidal radius $\sim 10^9$~cm, the resulting emission will be
extremely brief ($\sim 1$~sec), however if some mass is puffed out to
larger radii in the super-Eddington accretion phase of the merger, the
emission may be similar to that discussed here.  Interaction with a
spherical CSM is distinguishable from companion interaction by its
luminosity function; in the former case, the emission should be nearly
the same from all viewing angles.

In either case, the early time emission of supernovae provides much needed
insight into the nature of the progenitor system. Observational
surveys could be designed, either from space or the ground, to acquire
the collision signatures in a systematic way.  If one collects a
significant number of events, it will be possible to correlate the
measured separation distances with the properties of the ordinary \Nifs\ powered light curve
and spectra.  Such observations would
 provide direct, empirical insight into how the parameters of
the progenitor system influence the outcome of supernova
explosions.

\acknowledgements
The author is grateful for comments from E.~Ramirez-Ruiz, T.~Plewa,
 S.E~Woosley, and the referee C.~Matzner. 
Support  was provided by NASA through Hubble fellowship grant
\#HST-HF-01208.01-A awarded by the Space Telescope Science Institute,
which is operated by the Association of Universities for Research in
Astronomy, Inc., for NASA, under contract NAS 5-26555.  This research
has been supported by the DOE
SciDAC Program (DE-FC02-06ER41438).  Computing time was
provided by ORNL through an INCITE award and by NERSC.

\end{document}